\documentclass[preprint,12pt]{elsarticle}

\usepackage{amssymb}
\usepackage{amsmath}
\usepackage{siunitx}
\usepackage{subcaption}
\RequirePackage[subrefformat=parens]{subcaption}

\journal{Journal of Subatomic Particles and Cosmology}

\begin{document}

\begin{frontmatter}


\title{Probing $t$-channel single top-quark and antiquark production via differential cross-section measurements at $\sqrt{s}=$\SI{13}{\TeV} with the ATLAS detector}

\author[label1]{Lukas Kretschmann}
\author{on behalf of the ATLAS collaboration}
\affiliation[label1]{organization={University of Wuppertal},
             country={Germany}}

\begin{abstract}
The differential production cross-sections of single top quarks and top antiquarks produced via the $t$-channel process are measured in proton-proton collisions at $\sqrt{s}=\SI{13}{\TeV}$ at the LHC with the full Run~2 ATLAS dataset corresponding to an integrated luminosity of \SI{140}{\femto\barn^{-1}}. The cross-sections are measured as a function of the transverse momentum and absolute rapidity of the top quark ($tq$) and top antiquark ($\bar{t}q$) at parton level. In addition, for the first time, the differential ratio of the $tq$ to $\bar{t}q$ cross-sections is presented. The results are compared to theoretical predictions from fixed-order calculations, various event generators, and different PDF sets. An interpretation in the framework of an effective field theory (EFT) is performed to constrain the Wilson coefficient $C^{3,1}_{Qq}$ of the four-fermion operator.
\end{abstract}


\end{frontmatter}


\section{Introduction}
Single top-quark production via the $t$-channel exchange of a virtual $W$ boson is the dominant electroweak top-quark production process at the LHC. The process provides unique sensitivity to the structure of the proton and allows the study of the electroweak interaction of the top quark. The top-quark ($tq$) production cross-section is expected to be larger than the corresponding top-antiquark ($\bar{t}q$) cross-section due to the larger $u$-quark content in the proton compared to the $d$-quark content, making the ratio of them sensitive to PDF modelling~\cite{Campbell:2021qgd}.

The differential cross-section measurements target kinematic distributions such as the transverse momentum $p_\text{T}$ and the absolute rapidity $|y|$ of the produced top quark and antiquark. These observables are sensitive to the modelling of parton showers, matrix-element generators, and proton PDF sets. Furthermore, the ratio of $\sigma(tq)/\sigma(\bar{t}q)$ provides enhanced sensitivity to PDF uncertainties and benefits from significant cancellations of systematic uncertainties.

This work extends previous inclusive measurements by ATLAS~\cite{PERF-2007-01} by providing differential cross-sections at parton level and performing an EFT interpretation targeting the Wilson coefficient $C^{3,1}_{Qq}$ associated with the four-quark operator. This interpretation makes use of the dependence of selection efficiencies on the EFT contribution.

These proceedings are based on Ref.~\cite{ATLAS:2025vtm}.

\section{Analysis strategy and systematic uncertainties}
Events are selected to match the $t$-channel signature, requiring exactly one isolated charged lepton (electron or muon) with $p_\text{T} > 28$~GeV, exactly two jets with $p_\text{T} > 30$~GeV and $|\eta| < 4.5$ and exactly one $b$-tagged jet. The missing transverse energy is required to satisfy $E_\text{T}^\text{miss} > 30$~GeV, and the transverse mass of the $W$ boson must fulfill $m_\text{T}(W) > 50$~GeV.

To enhance signal purity, a feed-forward neural network (NN) is used to separate signal from background events. The NN assigns an output score $D_{\mathrm{nn}}$ to each event, with signal-like events receiving higher scores. A requirement of $D_{\mathrm{nn}} > 0.93$ is applied, corresponding to an optimised working point chosen to maximise the signal-to-background ratio while retaining sufficient signal statistics. This requirement yields $S/B = 6.1$ in the positive lepton signal region ($\ell^+$ SR) and $S/B = 3.8$ in the negative lepton region ($\ell^-$ SR). The two signal regions are defined based on the sign of the lepton charge to measure the cross-sections of $tq$ and $\bar{t}q$ separately.

The unfolded parton-level cross-sections are obtained by iterative Bayesian unfolding (IBU)~\cite{DAgostini:1994fjx}. The unfolding procedure corrects for detector effects, inefficiencies, and acceptance effects. Migration matrices are constructed using the nominal Monte Carlo (MC) signal predictions. Four iterations are chosen to minimise statistical fluctuations while reducing prior bias.

The unfolded differential cross-section is defined as:
\begin{equation*}
	\frac{\mathrm{d}\sigma_k}{\mathrm{d}X_k} = \frac{1}{\epsilon_k \mathcal{L}_{\mathrm{int}}\Delta X_k}\sum_j M^{-1}_{jk} (N^{\mathrm{data}}_j - \hat{B}_j),
\end{equation*}
where $M^{-1}_{jk}$ is the inverse of the migration matrix, $N^{\mathrm{data}}_j$ is the observed number of events, and $\hat{B}_j$ is the estimated background.

Systematic uncertainties are categorised into experimental, signal modelling, and background-related uncertainties. The dominant uncertainties arise from signal modelling (parton shower, scale variations, and PDF uncertainties) and experimental uncertainties such as jet energy scale and $b$-tagging efficiencies. Background uncertainties are subdominant due to the low background fraction in the signal regions.

\section{Results}
The unfolded parton-level differential cross-sections are measured as functions of $p_\text{T}(t)$, $p_\text{T}(\bar{t})$, $|y(t)|$, and $|y(\bar{t})|$. The ratio of $\sigma(tq)/\sigma(\bar{t}q)$ is also measured differentially. Unlike previous measurements~\cite{CMS:2019jjp}, which reported ratios of $\sigma(tq)$ to $\sigma(tq+\bar{t}q)$ due to limited statistics, this result provides, for the first time, a direct measurement of the $\sigma(tq)$ to $\sigma(\bar{t}q)$ ratio at differential level.

\begin{figure}[htb]
	\centering
	\begin{subfigure}{0.48\textwidth}
		\centering
		\includegraphics[width=\textwidth]{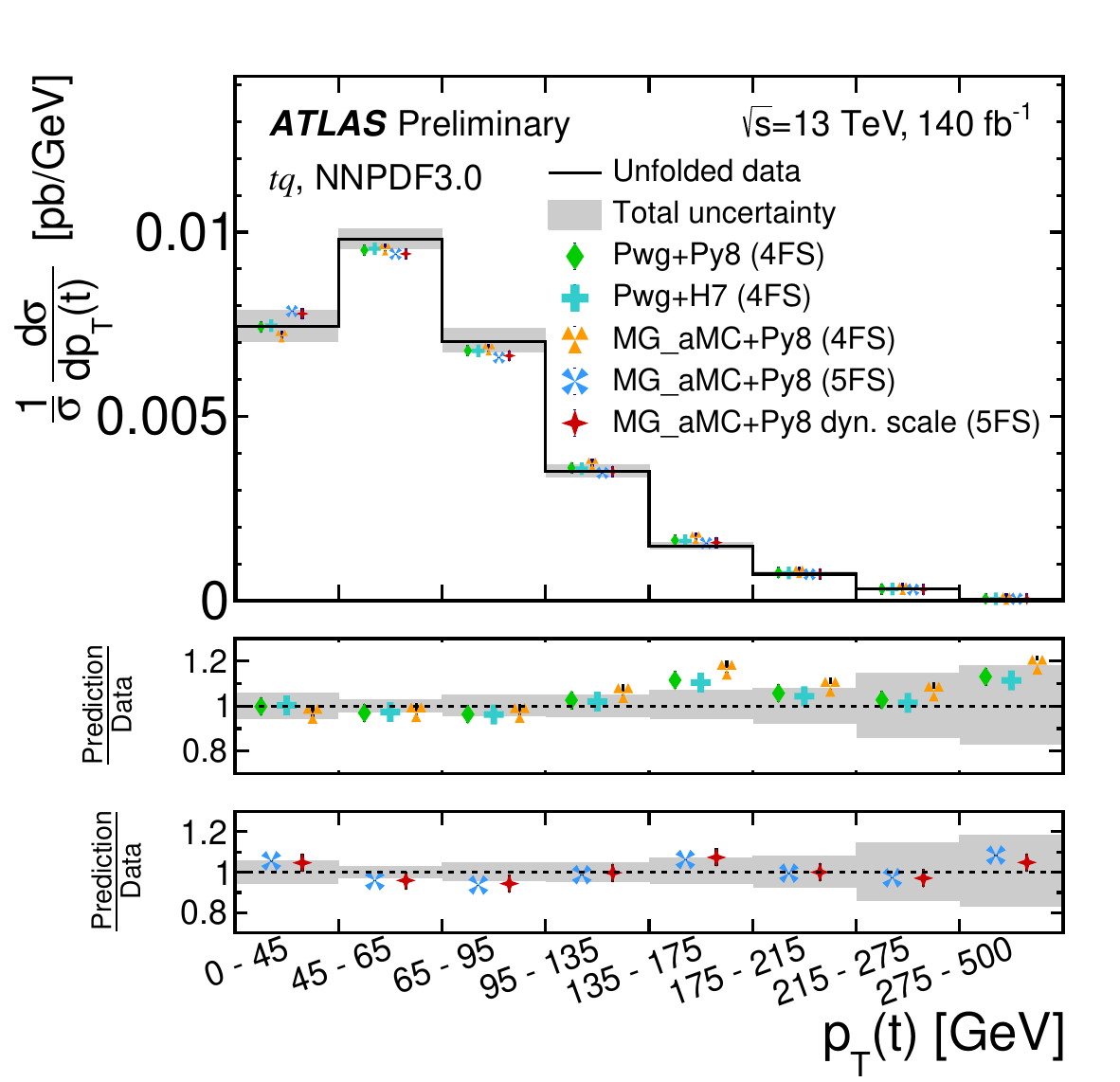}
		\caption{}
		\label{fig:pt_main}
	\end{subfigure}
	\begin{subfigure}{0.48\textwidth}
		\centering
		\raisebox{0.45cm}{
			\includegraphics[width=\textwidth]{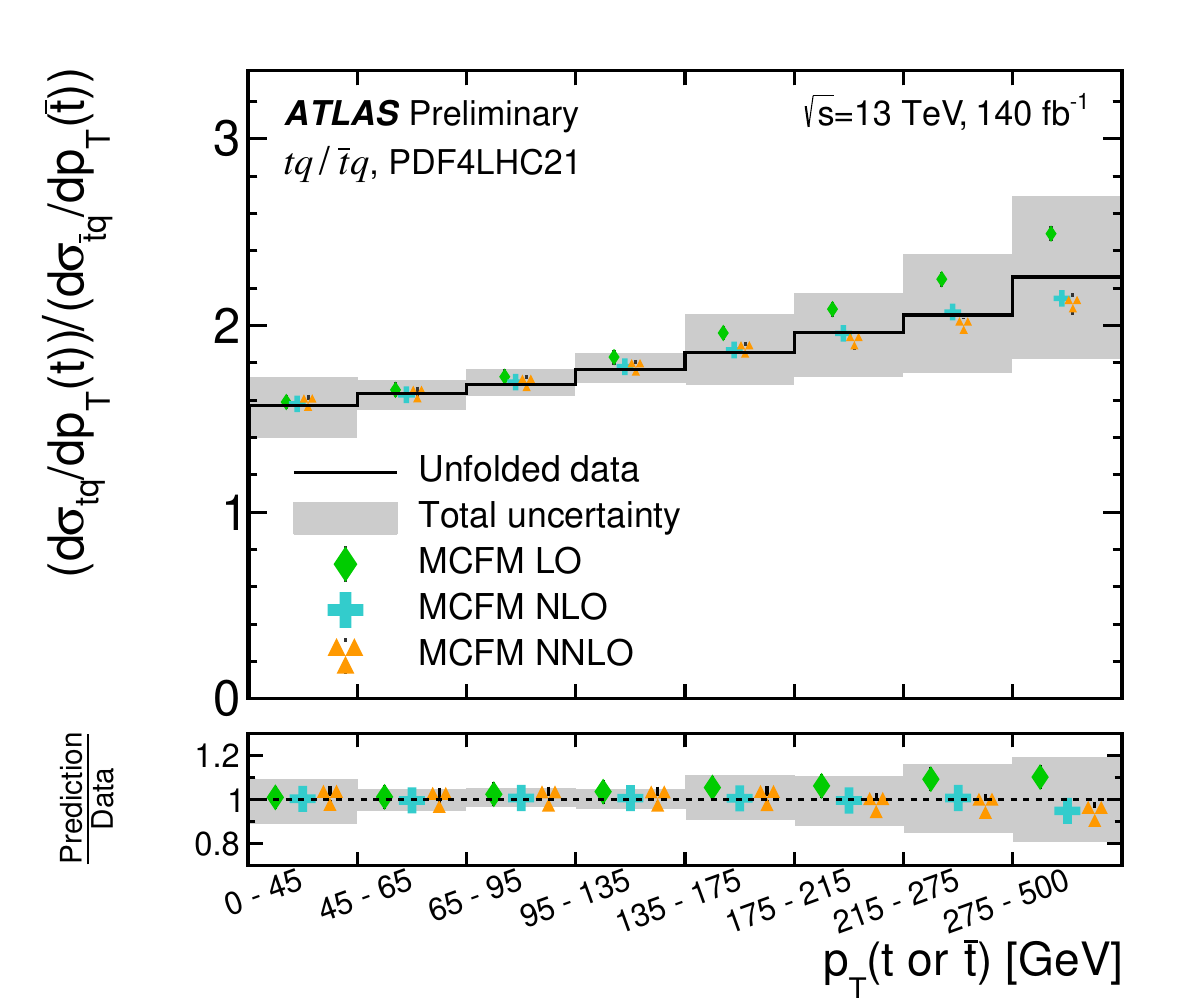}
		}
		\caption{}
		\label{fig:pt_ratio}
	\end{subfigure}
	
	\caption{\subref{fig:pt_main} The normalised differential $tq$ production cross-section as a function of $p_\text{T}(t)$ compared to theoretical predictions from different MC generators and \subref{fig:pt_ratio} the differential cross-section ratio as a function of $p_\text{T}(t)$ compared to fixed-order predictions with MCFM~\cite{Campbell:2020fhf}. The figures are both from~\cite{ATLAS:2025vtm}.}
	\label{fig:pt}
\end{figure}

\begin{figure}[htb]
	\centering
	\begin{subfigure}{0.48\textwidth}
		\centering
		\includegraphics[width=\textwidth]{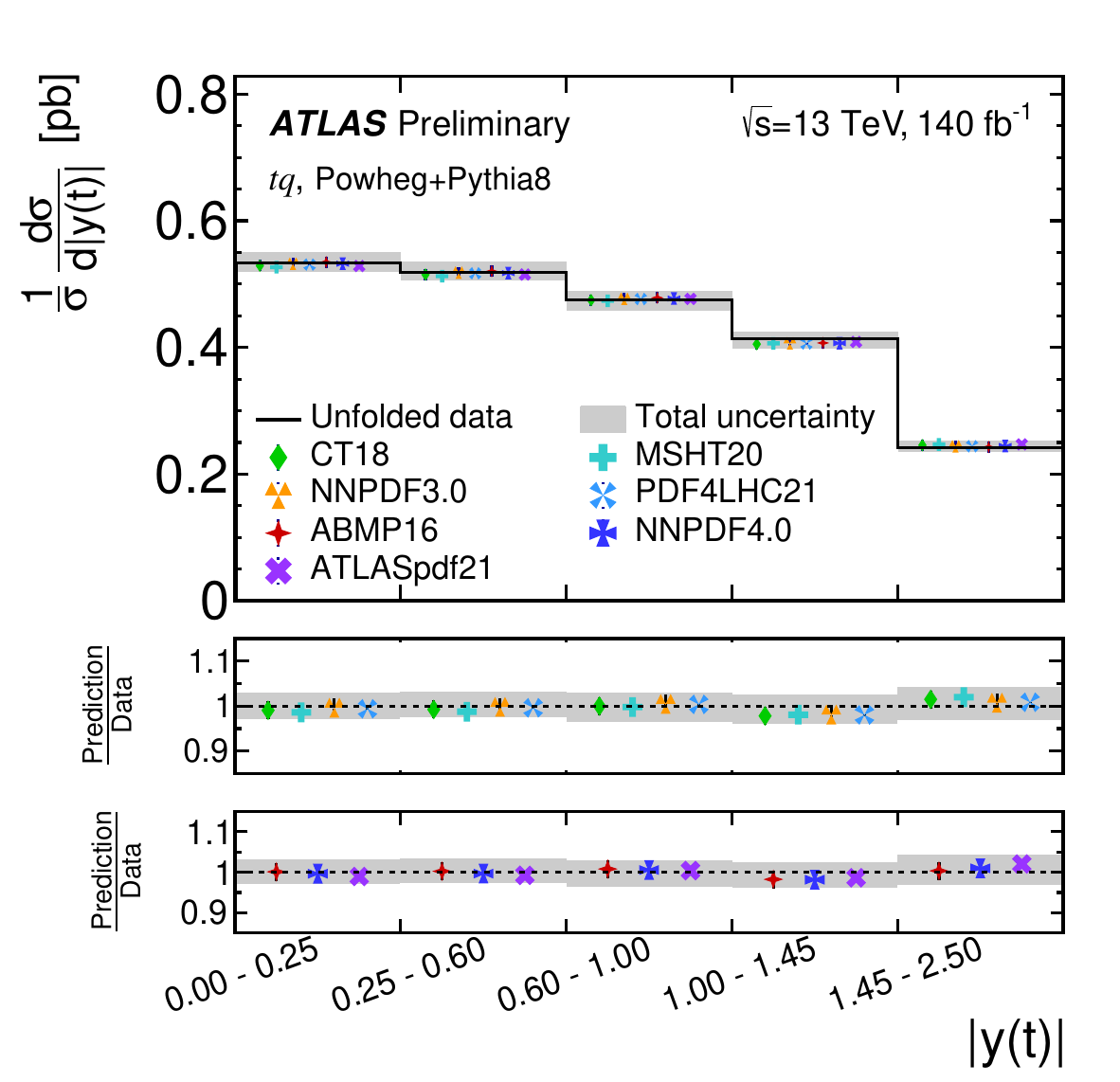}
		\caption{}
		\label{fig:y_main}
	\end{subfigure}
	\begin{subfigure}{0.48\textwidth}
		\centering
		\raisebox{0.45cm}{
			\includegraphics[width=\textwidth]{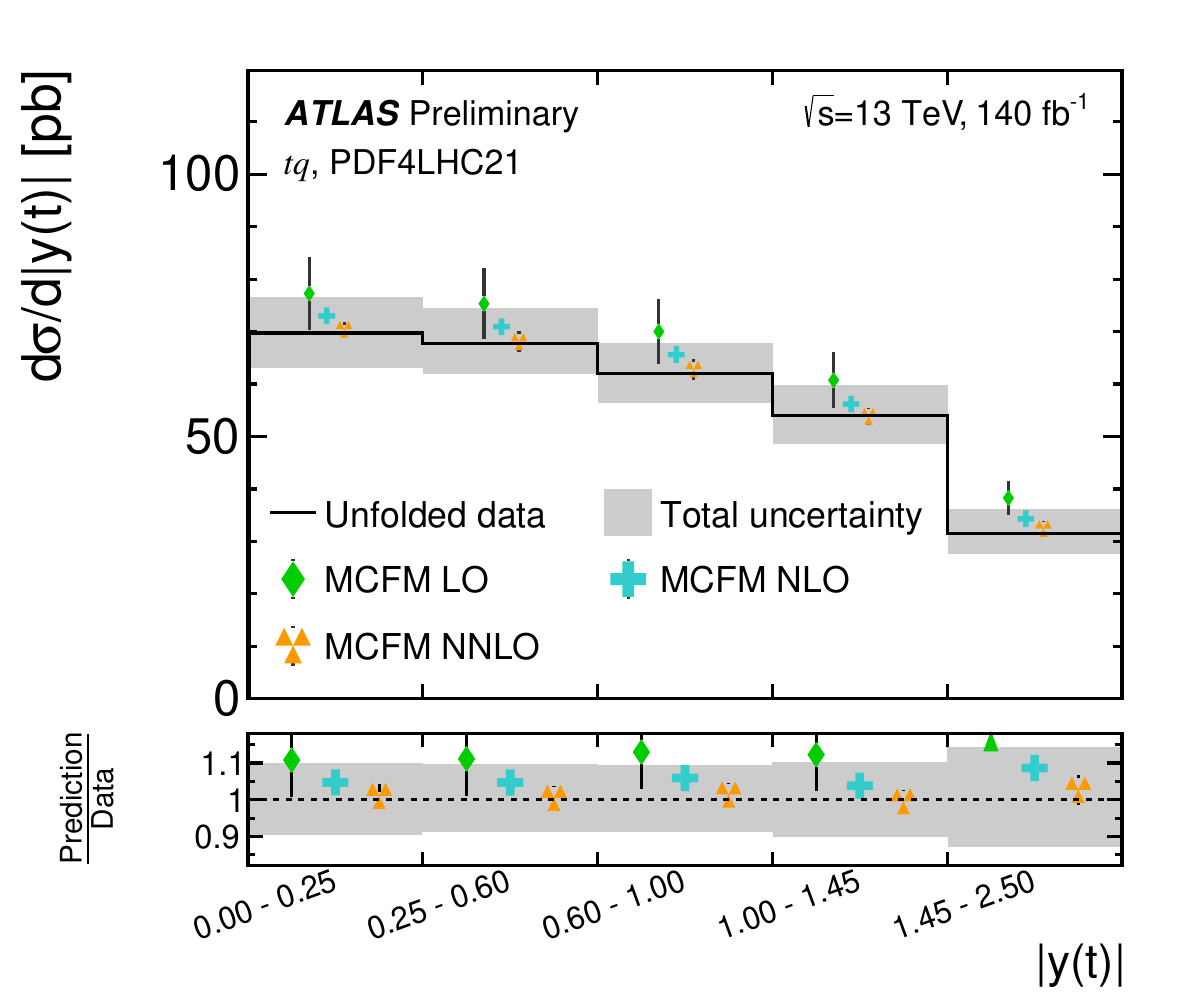}
		}
		\caption{}
		\label{fig:y_total}
	\end{subfigure}
	
	\caption{The \subref{fig:y_main} normalised differential $tq$ production cross-section as a function of $|y(t)|$ compared to theoretical predictions from different PDFs and \subref{fig:y_total} the differential cross-section as a function of $|y(t)|$ compared to fixed-order predictions with MCFM. The figures are both from~\cite{ATLAS:2025vtm}.}
	\label{fig:y}
\end{figure}

Good agreement is observed between data and theoretical predictions from Powheg+Pythia8, MG5\_aMC@NLO and MCFM~\cite{Campbell:2020fhf} at NNLO, as shown in Figure~\ref{fig:pt}. Figure~\ref{fig:y} also shows good agreement with different PDFs as well as LO to NNLO fixed-order prediction with MCFM~\cite{ATL-PHYS-PUB-2025-035}.

\section{EFT interpretation}
The measurement is interpreted in the Standard Model effective field theory (SMEFT) framework to constrain possible contributions from new physics via the operator $\mathcal{O}^{3,1}_{Qq}$~\cite{Buckley:2015lku}. Using detector-level samples with non-zero Wilson coefficients and unfolding to parton level, the dependency of the differential cross-sections on $C^{3,1}_{Qq}$ is parameterised as:

\begin{equation*}
	\frac{\mathrm{d}\sigma}{\mathrm{d}p_\text{T}(t)}(C^{3,1}_{Qq}) = \left(\frac{\mathrm{d}\sigma}{\mathrm{d}p_\text{T}(t)}\right)\cdot\left(1 + p_0\cdot C^{3,1}_{Qq} + p_1 \cdot \left(C^{3,1}_{Qq}\right)^2\right).
\end{equation*}

Using the EFTfitter tool~\cite{Castro:2016jjv}, the following constraint is obtained at 95\% confidence level:

\begin{equation*}
	-0.12~\mathrm{TeV}^{-2} < \frac{C^{3,1}_{Qq}}{\Lambda^2} < 0.12~\mathrm{TeV}^{-2},
\end{equation*}

which is a significant improvement over the results from the previous inclusive~\cite{ATLAS:2024ojr} analysis.

\section{Conclusion}
Differential $t$-channel single top-quark and top-antiquark production cross-sections are measured using the full ATLAS Run~2 dataset at the LHC. The results probe theoretical predictions, PDF modelling, and parton-shower configurations. The differential ratio $\sigma(tq)/\sigma(\bar{t}q)$ is measured for the first time. Overall, good agreement between all measured distributions and theoretical predicitons is observed. An EFT interpretation constrains the Wilson coefficient $C^{3,1}_{Qq}$ with improved precision over inclusive measurements.\\ \\
Copyright 2025 CERN for the benefit of the ATLAS Collaboration. Reproduction of this article or parts of it is allowed as specified in the CC-BY-4.0 license.

\bibliographystyle{elsarticle-num} 
\bibliography{references.bib}

@Article{PERF-2007-01,
	author         = "{ATLAS Collaboration}",
	title          = "{The ATLAS Experiment at the CERN Large Hadron Collider}",
	journal        = "JINST",
	volume         = "3",
	year           = "2008",
	pages          = "S08003",
	doi            = "10.1088/1748-0221/3/08/S08003",
	primaryClass   = "hep-ex",
}

@Booklet{ATLAS:2025vtm,
	author         = "{ATLAS Collaboration}",
	title = "{Measurement of differential $t$-channel single top (anti)quark production cross-sections at 13 TeV with the ATLAS detector}",	
	howpublished   = "{ATLAS-CONF-2025-011}",
	url            = "https://cds.cern.ch/record/2945445/",
	year           = "2025",
}

@article{DAgostini:1994fjx,
	author = "D'Agostini, G.",
	title = "{A Multidimensional unfolding method based on Bayes' theorem}",
	reportNumber = "DESY-94-099",
	doi = "10.1016/0168-9002(95)00274-X",
	journal = "Nucl. Instrum. Meth. A",
	volume = "362",
	pages = "487--498",
	year = "1995"
}

@article{Campbell:2020fhf,
	author = "Campbell, John and Neumann, Tobias and Sullivan, Zack",
	title = "{Single-top-quark production in the $t$-channel at NNLO}",
	eprint = "2012.01574",
	archivePrefix = "arXiv",
	primaryClass = "hep-ph",
	reportNumber = "FERMILAB-PUB-20-608-T, IIT-CAPP-20-05",
	doi = "10.1007/JHEP02(2021)040",
	journal = "JHEP",
	volume = "02",
	pages = "040",
	year = "2021"
}

@article{Campbell:2021qgd,
	author = "Campbell, John and Neumann, Tobias and Sullivan, Zack",
	title = "{Testing parton distribution functions with t-channel single-top-quark production}",
	eprint = "2109.10448",
	archivePrefix = "arXiv",
	primaryClass = "hep-ph",
	reportNumber = "FERMILAB-PUB-21-456-T, IIT-CAPP-21-01",
	doi = "10.1103/PhysRevD.104.094042",
	journal = "Phys. Rev. D",
	volume = "104",
	number = "9",
	pages = "094042",
	year = "2021"
}

@article{Buckley:2015lku,
	author = "Buckley, Andy and Englert, Christoph and Ferrando, James and Miller, David J. and Moore, Liam and Russell, Michael and White, Chris D.",
	title = "{Constraining top quark effective theory in the LHC Run II era}",
	eprint = "1512.03360",
	archivePrefix = "arXiv",
	primaryClass = "hep-ph",
	doi = "10.1007/JHEP04(2016)015",
	journal = "JHEP",
	volume = "04",
	pages = "015",
	year = "2016"
}

@article{Castro:2016jjv,
	author = {Castro, Nuno and Erdmann, Johannes and Grunwald, Cornelius and Kr{\"o}ninger, Kevin and Rosien, Nils-Arne},
	title = "{EFTfitter---A tool for interpreting measurements in the context of effective field theories}",
	eprint = "1605.05585",
	archivePrefix = "arXiv",
	primaryClass = "hep-ex",
	doi = "10.1140/epjc/s10052-016-4280-9",
	journal = "Eur. Phys. J. C",
	volume = "76",
	number = "8",
	pages = "432",
	year = "2016"
}

@Article{ATLAS:2024ojr,
	author = "{ATLAS Collaboration}",
	collaboration = "ATLAS Collaboration",
	title = "{Measurement of t-channel production of single top quarks and antiquarks in pp collisions at 13 TeV using the full ATLAS Run 2 data sample}",
	eprint = "2403.02126",
	archivePrefix = "arXiv",
	primaryClass = "hep-ex",
	reportNumber = "CERN-EP-2023-289",
	doi = "10.1007/JHEP05(2024)305",
	journal = "JHEP",
	volume = "05",
	pages = "305",
	year = "2024",
	note = "[Erratum: JHEP 06, 024 (2024)]"
}

@Article{ATL-PHYS-PUB-2025-035,
	author = "{ATLAS and CMS Collaborations}",
	title         = "{Reference Single Top-Quark Cross-Sections for ATLAS and
	CMS Analyses}",
	institution   = "CERN",
	reportNumber  = "ATL-PHYS-PUB-2025-035",
	address       = "Geneva",
	year          = "2025",
	url           = "https://cds.cern.ch/record/2942746",
}

@article{CMS:2019jjp,
	author = "Sirunyan, Albert M and others",
	collaboration = "CMS",
	title = "{Measurement of differential cross sections and charge ratios for t-channel single top quark production in proton{\textendash}proton collisions at $\sqrt{s}=13\,\text {Te}\text {V}$}",
	eprint = "1907.08330",
	archivePrefix = "arXiv",
	primaryClass = "hep-ex",
	reportNumber = "CMS-TOP-17-023, CERN-EP-2019-138",
	doi = "10.1140/epjc/s10052-020-7858-1",
	journal = "Eur. Phys. J. C",
	volume = "80",
	number = "5",
	pages = "370",
	year = "2020"
}

\end{document}